\documentclass[%
nofootinbib,
 amsmath,amssymb,
 aps,
prd,showpacs,twocolumn,superscriptaddress
]{revtex4-2}
\usepackage{graphicx}
\usepackage{subfigure}
\usepackage{dcolumn}
\usepackage{tabularx}
\usepackage{booktabs}
\usepackage{amsmath}
\newtheorem{theorem}{Theorem}
\usepackage{amssymb}
\usepackage{bm}
\usepackage{color}

\def\ttau{\tilde{\tau}}
\def\tlam{\tilde{\lambda}}

\def\tf{\tilde{f}}
\def\tx{\tilde{x}}
\usepackage[normalem]{ulem}
\usepackage[dvipsnames]{xcolor}
\usepackage{bm}
\usepackage{hyperref}

\begin{document}

\preprint{APS/123-QED}

\title{A New High-Performing Method for Solving the Homogeneous Teukolsky Equation}

\author{Ye Jiang}
 \affiliation{Shanghai Astronomical Observatory, Shanghai, 200030, China}
\affiliation{School of Astronomy and Space Science, University of Chinese Academy of Sciences,
Beijing, 100049, China}
\author{Wen-Biao Han}%
 \email{wbhan@shao.ac.cn}
\affiliation{Shanghai Astronomical Observatory, Shanghai, 200030, China}
\affiliation{School of Fundamental Physics and Mathematical Sciences, Hangzhou Institute for Advanced Study, UCAS, Hangzhou 310024, China}
\affiliation{School of Astronomy and Space Science, University of Chinese Academy of Sciences,
Beijing, 100049, China}
\affiliation{Taiji Laboratory for Gravitational Wave Universe (Beijing/Hangzhou), University of Chinese Academy of Sciences, Beijing 100049, China;}
\affiliation{Key Laboratory of Radio Astronomy and Technology (Chinese Academy of Sciences), A20 Datun Road, Chaoyang District, Beijing, 100101, China}
\date{\today}

\begin{abstract}
The numerical waveforms for the extreme mass-ratio inspirals (EMRIs) require a huge amount of homogeneous solutions of the Teukolsky equation in the frequency domain. The calculation accuracy and efficiency of the homogeneous solutions are the key performance bottleneck in waveform generation. In this paper, we propose a new numerical method based on the analytical series expansion which is most efficient for computing the homogeneous solutions with very high accuracy and a wider frequency range compared with the existing methods. Our new method is definitely useful for constructing the waveform templates of EMRIs.
\end{abstract}

\maketitle


\section{Introduction}\label{sec: Introduction}
The landmark detection of gravitational waves (GWs) by LIGO/Virgo collaborations \cite{theligoscientificcollaboration2021gwtc3} has ushered in a new era of multimessenger astronomy, where the ripples of spacetime curvature encode vital information about compact binary coalescences \cite{PhysRevLett.116.061102}. In addition to the ground-based detections, which focus on the frequency $10^1\sim10^3\;\text{Hz}$, the space-borne detections will focus on the frequency $10^{-5}\sim10^{0}\;\text{Hz}$, e.g., Taiji \cite{10.1093/nsr/nwx116}, LISA \cite{amaroseoane2017laser}, and Tian-Qin \cite{Luo_2016}. These space-borne detectors can capture GWs emanating from a diverse array of astrophysical and cosmological sources \cite{Amaro-Seoane2023}.

Extreme mass-ratio inspirals are one of the most interesting GW sources for upcoming space-borne GW detectors. In an EMRI, a stellar secondary of mass $\mu$ evolves around a supermassive black hole (SMBH) of mass $M$ with mass ratio $q=\mu/M\sim 10^{-4}-10^{-7}$, and the secondary object completes approximately $\mathcal{O}(1/q)$ orbits around the primary before its final plunge \cite{PhysRevD.95.103012}. Because of the long-time evolution of EMRIs, GWs produced by this kind of system will carry rich information about the strong gravitational field around the SMBHs. Thus, EMRIs are expected to offer unprecedented precision in parameter estimation for astrophysics \cite{AmaroSeoane2021,10.1093/mnras/stab2741,PhysRevLett.129.241103} and fundamental physics \cite{Barack_2019,Barausse2020}.

However, in the actual utilization of the detection, the GW signal is mounted on a huge noise and requires accurate waveform templates to extract it from the noisy data \cite{PhysRevD.104.064047}. The primary method of modeling EMRIs is the black hole perturbation theory \cite{Pound2020}, which solves the Einstein field equations by separating the full metric into the background part and perturbation one. During the inspiral, the motion of the stellar compact object generates persistent gravitational radiation. Thus, EMRIs can be modeled as a perturbation to the Kerr background generated by SMBHs.

Teukolsky had done a detailed study of the perturbation with the Kerr background \cite{1973ApJ...185..635T,1973ApJ...185..649P,1974ApJ...193..443T}. In the Teukolsky formalism, the differential equations of several scalars are given instead of directly solving the Einstein field equations; these scalars encode the radiation information for various fields. The Teukolsky equation is written as \cite{1973ApJ...185..635T}
\begin{equation}
\begin{aligned}
&\left[\frac{(r^2+a^2)^2}{\Delta}-a^2\sin^2{\theta}\right]\frac{\partial^2\psi}{\partial t^2}\\
&+\frac{4Mar}{\Delta}\frac{\partial^2\phi}{\partial t\partial \phi}+\left[\frac{a^2}{\Delta}-\sin^{-2}{\theta}\right]\frac{\partial^2\psi}{\partial\phi^2}\\
&-\Delta^{-s}\frac{\partial}{\partial r}\left(\Delta^{s+1}\frac{\partial\psi}{\partial r}\right)-\sin^{-1}{\theta}\frac{\partial}{\partial\theta}\left(\sin{\theta}\frac{\partial\psi}{\partial\theta}\right)\\
&-2s\left[\frac{a(r-M)}{\Delta}+\frac{i\cos{\theta}}{\sin^2{\theta}}\right]\frac{\partial\psi}{\partial\phi}\\
&-2s\left[\frac{M(r^2-a^2)}{\Delta}-r-ia\cos{\theta}\right]\frac{\partial\psi}{\partial t}\\
&+(s^2\cot^2{\theta}-s)\psi=4\pi\Sigma T
\end{aligned}
\end{equation}
where $\Delta=r^2+a^2-2r$, $\Sigma=r^2+a^2\cos{\theta}$, $T$ is the source term for the Teukolsky equation, and $\psi$ can related to different scalars for different spin weights $s$. Here, $\psi$ with $s=0,\pm1,\pm2$ represent the scalar, electromagnetic, and gravitational radiation, respectively. Moreover, after transforming the equation to the frequency domain, the Teukolsky equation can be separated into two parts, the radial part of which is given by
\begin{equation}\label{eq: Teukolsky radial equation}
\begin{aligned}
&\Delta^{-s}\frac{d}{dr}\left(\Delta^{s+1}\frac{dR}{dr}\right)\\
&-\left(\lambda-4is\omega r-\frac{K^2-2is(r-M)K}{\Delta}\right)R=\mathcal{T}
\end{aligned}
\end{equation}
where $K=(r^2+a^2)\omega-ma$, $\mathcal{T}$ is the source term projected onto the angular part eigenfunction, and $\lambda$ is the separation constant or the eigenvalue of the angular part. Our main text will focus on the radial part, and the discussion of the eigenvalue and eigenfunction of the angular part can be found in Appendix \ref{app: Angular Teukolsky Equation}. In Ref.~\cite{PhysRevD.73.024027}, Eq.~(\ref{eq: Teukolsky radial equation}) is solved using the Green function method, in which the Green function is constructed of homogeneous solutions for Eq.~(\ref{eq: Teukolsky radial equation}). Overall, a huge amount of homogeneous solutions are required in the numerical calculation of the Teukolsky-based waveforms for EMRIs. Therefore, the computational efficiency of the homogeneous solutions is crucial in the EMRI waveform calculation.

Previous studies have made many efforts to increase the calculation efficiency of homogeneous solutions. Leaver formulated a method to express them in terms of a series of Coulomb wave functions \cite{10.1063/1.527130}. However, this series converges only when $r$ is not close to the horizon. Thus, Mano, Suzuki and Takasugi (MST) constructed a method that adopts the hypergeometric functions to represent the homogeneous solutions and converges around the horizon \cite{10.1143/PTP.95.1079}. With these formulas, it is possible to match the two series analytically and then obtain the asymptotic amplitude \cite{10.1143/PTP.112.415,10.1143/PTP.113.1165}. Although the MST method can theoretically solve the homogeneous equation at an arbitrary frequency, the performance will be poor when $\omega$ becomes large.

The simplest way to solve a differential equation is the direct integration method. However, note that in Eq.~(\ref{eq: Teukolsky radial equation}), the potential falls off as $1/r$ when $r$ reaches infinity. This long-range potential makes numerical integration more difficult. Another method, the generalized Sasaki-Nakamura (SN) formalism \cite{SASAKI198185,SASAKI198268,10.1143/PTP.67.1788,PhysRevD.62.044029}, can alleviate this phenomenon by numerical integration. To overcome this difficulty, the SN method first transforms the potential to the well-behaved one ($\sim r^{-2}$), and then performs numerical integration. A detailed and systematic introduction can be found in Ref.~\cite{PhysRevD.110.124070}. However, the numerical integration can encounter an oscillation problem at large $\omega$, which will decrease the efficiency. Recently, numerical methods using the confluent Heun function have also been developed with high accuracy \cite{Chen2024}.

In this paper, we describe a new method (hereafter the Jiang-Han or JH method) based on the power series expansion at ordinary points and various asymptotic series expansions at singular points. The JH method replaces the numerical integration with analytic expansion and matches the asymptotic series expansion to obtain the asymptotic amplitudes. In detail, we drive the recursion for the coefficients of each series expansion and discuss the asymptotic behavior for those difference equations by Poincar$\acute{\text{e}}$-Perron theory. Furthermore, for large $\omega$ cases, we introduce the large frequency expansion to solve the oscillation problem, which induces inaccurate results and slows down the convergence speed.

This paper is organized as follows. In Sec.~\ref{sec: Analytical Series Expansion Method}, we first review the basic properties of the Teukolsky radial equation, including asymptotic behavior and singularity. Then, we introduce our series expansion in detail. For large $\omega$, we provide a large value expansion in Sec.~\ref{sec: Large Frequency Expansion}. In Sec.~\ref{sec: Numerical Experiment}, we perform some numerical experiments and compare them with the original MST and SN methods. Finally, we summarize our results in Sec.~\ref{sec: Conclusion}. Throughout this paper, we adopt the geometric units $G=c=1$ and assume $M=1$.

\section{Analytical Series Expansion Method}\label{sec: Analytical Series Expansion Method}
We first rewrite the radial Teukolsky equation as follows:
\begin{equation}\label{eq: minor changed teukolsk equation}
R^{\prime\prime}(r)+\frac{2(s+1)(r-1)}{\Delta}R^\prime(r)-\frac{V(r)}{\Delta}R(r)=0
\end{equation}
with
\begin{equation}\label{eq: teukolsky potential}
V(r)=\lambda-4is\omega r-\frac{K^2-2is(r-1)K}{\Delta}.
\end{equation}
Now, considering the analyticity of the coefficient in front of $R^\prime(r)$, $R(r)$, one can find that the Teukolsky equation has two regular singularities at $r=r_\pm=1\pm\sqrt{1-a^2}$, and an irregular singularity point at $r=\infty$ with rank 1. With this information, the leading order of the asymptotic solutions at $r\rightarrow r_+,\infty$ are
\begin{align}
R(r\rightarrow r_+)&\sim C_1\Delta^{-s}e^{-ipr_*}+C_2e^{ipr_*},\\
R(r\rightarrow \infty)&\sim C_3r^{-1}e^{-i\omega r_*}+C_4r^{-2s-1}e^{i\omega r_*},
\end{align}
with
\begin{equation}
r_*=r+\frac{2r_+}{r_+-r_-}\ln{\left(\frac{r-r_+}{2}\right)}+\frac{2r_-}{r_--r_+}\ln{\left(\frac{r-r_-}{2}\right)},
\end{equation}
where $p=\omega-ma/r_+$ and $C_{1\sim4}$ are constants. Furthermore, through this paper, we will assume $\omega\neq 0$ and $p\neq 0$. In physical applications, to force the Green function to satisfy the ingoing boundary condition at the horizon and the outgoing boundary condition at infinity concurrently, we select two classes of solutions $R^{\text{in}}$ and $R^{\text{up}}$, which satisfy the ingoing boundary condition at the horizon and the outgoing boundary condition at infinity, respectively (not concurrently). In the explicit expression, they have the following properties:
\begin{equation}\label{eq: rin bound condition}
R^{\text{in}}(r)=\left\{\begin{aligned}
&B^{\text{trans}}\Delta^{-s}e^{-ipr_*},&r\rightarrow r_+\\
&B^{\text{inc}}r^{-1}e^{-i\omega r_*}+B^{\text{ref}}r^{-2s-1}e^{i\omega r_*},&r\rightarrow \infty
\end{aligned} \right.
\end{equation}
\begin{equation}\label{eq: rup bound condition}
R^{\text{up}}(r)=\left\{\begin{aligned}
&C^{\text{ref}}\Delta^{-s}e^{-ipr_*}+C^{\text{inc}}e^{ipr_*},&r\rightarrow r_+\\
&C^{\text{trans}}r^{-2s-1}e^{i\omega r_*},&r\rightarrow \infty
\end{aligned} \right.
\end{equation}
In this study, we focus on the numerical calculation of these two classes of solutions and their asymptotic coefficients. To simplify the discussion, we first change the position of the singularities at $r_-$, $r_+$ to $1$ and $0$, respectively, by setting $x=(r_+-r)/(r_+-r_-)$; then, we perform the transformation $R(r)=T(x)P(x)$, this step is aim to reduce the complexity of the series expansion in the following subsections, where $T(x)$ takes the following form:
\begin{equation}
T(x)=e^{i\epsilon\kappa x}(-x)^{-s-i(\epsilon+\tau)/2}(1-x)^{i(\epsilon-\tau)/2},
\end{equation}
where $\epsilon=2\omega$, $\kappa=\sqrt{1-a^2}$, and $\tau=(\epsilon-ma)/\kappa$.
After this transformation, $P(x)$ should satisfy:
\begin{align}
x(1-x)&\frac{d^2P(x)}{dx^2}+q_1(x)\frac{dP(x)}{dx}+q_2(x)P(x)=0,\label{eq: P(x)}\\
q_1(x)=&(1-s-i\epsilon-i\tau)-2(1-i\epsilon\kappa-i\tau)x-2i\epsilon\kappa x^2,\\
q_2(x)=&\lambda-\epsilon^2+i\epsilon\kappa(1-2s)+s(s+1)+\tau(i+\tau)\nonumber\\
&+2\epsilon\kappa(-i+is+\epsilon-\tau)x.
\end{align}
Different from the MST method, which provides a global solution in terms of hypergeometric functions, our method provides various types of local solutions that are only valid in a finite range and finally packages them into a global solution of $P(x)$.

For the convenience of discussing the convergence property of the local solution, we introduce the Poincar$\acute{\text{e}}$-Perron theory here. The proof and further discussion can be found in Ref.~\cite{Elaydi2005}.
\begin{theorem}\label{theorem: 1}
let $\{x_n\}_{n=0}^{\infty}$ be a sequence generated by the recursion relation
\begin{equation}
\left\{ \begin{array}{ll}
&x_{n+1}=A_nx_n+B_nx_{n-1}\\
&x_0=1
\end{array}\right.
\end{equation}
where $n\in\mathbb{Z}^+$the coefficients have well-defined limits as $n\rightarrow\infty$,
\begin{equation}
\left\{
\begin{array}{ll}
&\lim_{n\rightarrow\infty}A_n=A\\
&\lim_{n\rightarrow\infty}B_n=B
\end{array}\right.
\end{equation}
The roots of the characteristic equation $\lambda^2=A\lambda+B$ are denoted $\lambda=\lambda_1,\lambda_2$. We assume that distinct characteristic roots have distinct moduli $|\lambda_1|<|\lambda_2|$.
Then there exists a fundamental set of solutions, $\{a_n\}_{n=0}^{\infty}$ and $\{b_n\}_{n=0}^{\infty}$ which satisfy
\begin{equation}
\left\{
\begin{array}{ll}
&\lim_{n\rightarrow\infty}\frac{a_{n+1}}{a_n}=\lambda_1,\\
&\lim_{n\rightarrow\infty}\frac{b_{n+1}}{b_n}=\lambda_2,
\end{array}\right.
\end{equation}
and every solution of the recursion relation can be represented by a linear combination of the elements in this fundamental set.
\end{theorem}
From this theorem, for any initial condition $x_1$,
\begin{equation}
\lim_{n\rightarrow\infty}\frac{x_{n+1}}{x_n}=\lambda_2,
\end{equation}
except $x_1$ is exactly equal to $b_1$.

\subsection{Local Solution at Zero}\label{sec: Horizon Solution}
We first discuss the local solution at 0, which we denote as $P_0$. Note that $x=0$ corresponds to the horizon $r_+$, thus, this solution also reflects the asymptotic behavior of the homogeneous solutions around the horizon. For Eq.~(\ref{eq: P(x)}), $x=0$ is a regular singularity; thus, the two independent solutions can be derived by the Frobenius method \cite{NIST:DLMF}. The results read
\begin{align}
P_{0}^{\text{in}}(x)&=\sum_{n=0}^{\infty} a_n x^n,\label{eq: local horizon in}\\
P_{0}^{\text{out}}(x)&=(-x)^{s+i(\epsilon+\tau)}\sum_{n=0}^{\infty} b_n x^n,\label{eq: local horizon out}
\end{align}
where $P_{0}^{\text{in}}$ and $P_{0}^{\text{out}}$ are the solutions that correspond to the ingoing and outgoing boundary conditions, respectively. $a_n$ and $b_n$ can be obtained from
\begin{widetext}
\begin{align}
a_n=&\frac{ -\lambda +(n-s-i \tau -1) (n+s-i \tau )+i \kappa  \epsilon  (-2 n+2 s+1)+\epsilon ^2}{n (n-s-i(\tau +\epsilon) )}a_{n-1}+\frac{2 i \kappa  \epsilon  ( (n-s-1)-i(\tau -\epsilon) )}{n (n-s-i(\tau +\epsilon))}a_{n-2},\\
b_n=&\frac{-\lambda +i \epsilon  (\kappa -2 (\kappa -1) n+2 s-1)+(n-1) (n+2 s)+2 \kappa  \epsilon ^2+2 \kappa  \tau  \epsilon }{n (n+s+i (\tau +\epsilon ))}b_{n-1}+\frac{2 i \kappa  \epsilon  (n+2 i \epsilon -1)}{n (n+s+i (\tau +\epsilon ))}b_{n-2}.
\end{align}
\end{widetext}
with $a_0=b_0=1$ and $a_{-1}=b_{-1}=0$.

The convergence radii of the series are not larger than the distance of the nearest singularity from $0$, which can also be obtained by Theorem \ref{theorem: 1}
\begin{equation}
\lim_{n\rightarrow \infty}\frac{a_{n+1}}{a_n}=\lim_{n\rightarrow \infty}\frac{b_{n+1}}{b_n}=1.
\end{equation}
It announces that Eq.~(\ref{eq: local horizon in}), Eq.~(\ref{eq: local horizon out}) are only valid when $|x|<1$. However, $a_n$ frequently converge to a pretty large number for a large $|\omega|$ and results in a polynomial with a large condition number. Thus, in practice, we truncate the series $a_n$ at $N=20$ to reduce the error and set the convergence range $\mathcal{R}$ as
\begin{equation}\label{eq: coverage range}
\mathcal{R}=\left(\frac{\bf{tol}}{|a_N|}\right)^{\frac{1}{N}}
\end{equation}
for some tolerance $\bf{tol}$.

On the contrary, $\{b_n\}_{n=0}^\infty$ always converges to a tiny number. Consequentially, if we directly truncate $\{b_n\}_{n=0}^\infty$ at some $n$ and use the Eq.~(\ref{eq: coverage range}), it will raise a result much larger than $1$. Thus, the following expression is used:
\begin{equation}
\mathcal{R}=\min\left\{\left(\frac{\bf{tol}}{|b_N|}\right)^{\frac{1}{N}},0.6\right\}
\end{equation}

Through the discussion above, we have assumed that $-s-i(\epsilon+\tau)$ is not an integer number. Otherwise, it will introduce an extra term proportional to $\ln{x}$ into the solution, see also Eq.~(2.7.6) in Ref.~\cite{NIST:DLMF}. For an integer $s$, it is equal to say $\omega=m a/(2 r_+)$, which relates to the superradiance.

\subsection{Local Solution at Infinity}\label{sec: Infinity Solution}
We then discuss the local solution at infinity. The point $x=-\infty$ corresponds to the irregular rank 1 singular point $r=\infty$ in Eq.~(\ref{eq: P(x)}). Hence, there exist two formal asymptotic expansions, which are
\begin{align}
P_{\infty}^{\text{in}}(x)=&(-x)^{-i \epsilon+s+i \tau-1}\sum_{n=0}^{\infty}a_n\frac{1}{x^n}\label{eq: inf expand 1}\\
P_{\infty}^{\text{out}}(x)=&e^{-2 i \epsilon \kappa x} (-x)^{i \epsilon-s+i \tau-1}\sum_{n=0}^{\infty}b_n\frac{1}{x^n} \label{eq: inf expand 2}
\end{align}
where $P_{\infty}^{\text{in}}$ and $P_{\infty}^{\text{out}}$ are the solutions corresponding to the ingoing and outgoing boundary conditions, respectively. $a_n$ and $b_n$ can be obtained from the following three-term recursion relation:
\begin{widetext}
\begin{align}
a_n=&\frac{(n+i \epsilon ) (n+2 i \kappa  \epsilon +i \epsilon-2 s -1)-\lambda +\epsilon ^2+\kappa  (2 \tau -i) \epsilon}{2 i \epsilon\kappa  n  }a_{n-1}+\frac{(n+2 i \epsilon -1) (i\tau - (n-s+i \epsilon -1))}{2 i \epsilon \kappa  n  }a_{n-2},\\
b_n=&\frac{-i \lambda +i n^2+n (2 i s+2 (\kappa +1) \epsilon -i)+2 s (\kappa  \epsilon +\epsilon -i)-(\kappa +1) \epsilon }{2 \kappa  n \epsilon }b_{n-1}+\frac{(n+2 s-1) (n+s-i (\tau +\epsilon -i))}{2 i \kappa  n \epsilon }b_{n-2}.
\end{align}
\end{widetext}
with $a_0=b_0=1$ and $a_{-1}=b_{-1}=0$.

It can be found that both of the recurrence relations are divergent with
\begin{equation}
\lim_{n\rightarrow \infty} \frac{a_n}{n a_{n-1}}=\lim_{n\rightarrow \infty} \frac{b_n}{n b_{n-1}}=\frac{1}{2i\epsilon\kappa}.
\end{equation}
Consequently, it is impossible to bound the tail by an equation similar to Eq.~(\ref{eq: coverage range}). To obtain an accurate result using formal asymptotic expansion, we must choose a ``suitable" large value of $x$ to minimize the influence of the remainder term. The following expression is a good choice:
\begin{equation}
\mathcal{R}=\left|\frac{a_N}{\bf{tol}}\right|^\frac{1}{N},
\end{equation}
where $N$ is choose to minimized $\mathcal{R}$, we also truncate the asymptotic series at $n=N$ and assume the asymptotic series are valid when $|x|>\mathcal{R}$.Additionally, from the discussion in Ref.~\cite{doi:10.1137/1038006}, if the truncated asymptotic series is correct for certain values of $x_0$, then it should also be correct for $|x|>|x_0|$. Hence, it is better to perform a consistency check at $x=-\mathcal{R}$ to ensure accuracy. The details are described in Appendix \ref{app: Check the Asymptotic Expansion at Infinity}.

There definitely exist other methods to bound the remainder and increase the accuracy of the asymptotic series, such as exponentially improved expansions \cite{doi:10.1098/rspa.1994.0047} or sequence transformations \cite{WENIGER1989189}. Here, we choose the simplest way to overcome this difficulty.

In practice, $\mathcal{R}$ will become larger when $\omega$ close to 0, that because our asymptotic series Eq.~(\ref{eq: inf expand 1}) and Eq.~(\ref{eq: inf expand 2}) developed based on the irregular rank 1 singular point at $x=-\infty$. When $\omega\rightarrow0$, this irregular singular point will transform into a regular singular point; thus, our assumption is broken. Consequently, the asymptotic series is invalid, and a single hypergeometric function can represent the whole function. Thus, following Ref.~\cite{10.1063/1.527130}, we expand the solution into a series of confluent hypergeometric functions; details can be found in Appendix \ref{app: Confluent Hypergeometric Expansion}.

\subsection{Ordinary Point Expansion}\label{sec: Ordinary  Point Expansion}
With the asymptotic solution at $0$ and $-\infty$ in hand, we still need a ``bridge" to connect those solutions and form a global solution finally. For any analytic function $P(x)$ in the domain $|x-x_0|<\mathcal{R}$, we have the following series expansion:
\begin{equation}
P(x)=P(x_0)\sum_{n=0}^{\infty}a_n(x-x_0)^n.
\end{equation}
By substituting this ordinary point series expansion into the Eq.~(\ref{eq: P(x)}), the following recurrence relation can be obtained by equating the coefficients of $x^n$:
\begin{widetext}
\begin{align}\label{eq: ordinary}
a_n=&-\frac{2 x_0 (n-i \tau -1)-n+s+i \tau +i \epsilon  \left(2 \kappa  \left(x_0-1\right) x_0+1\right)+1}{n \left(x_0-1\right) x_0}a_{n-1}+\frac{2 \kappa  \epsilon  (-i (n-s-2)-\tau +\epsilon )}{(n-1) n \left(x_0-1\right) x_0}a_{n-3}\\\nonumber
&+\frac{-\lambda +(n-s-i \tau -2) (n+s-i \tau -1)+i \kappa  \epsilon  \left(x_0 (4 n-2 (s+i \tau +3))-2 n+2 s+3\right)+\epsilon ^2 \left(1-2 \kappa  x_0\right)}{(n-1) n \left(x_0-1\right) x_0}a_{n-2},
\end{align}
\end{widetext}
with $a_1=P^\prime(x_0)/P(x_0)$, $a_{0}=1$ and $a_{-1}=0$.

For a large $n$, Eq.~(\ref{eq: ordinary}) will become
\begin{equation}
a_{n}=a_{n-1}\frac{1-2x_0}{x_0(x_0-1)}-a_{n-2}\frac{1}{x_0(x_0-1)}.
\end{equation}
Thus, by Theorem \ref{theorem: 1}, we have
\begin{equation}\label{eq: coverage ordinary}
\lim_{n\rightarrow\infty}\frac{a_n}{a_{n-1}}=-\frac{1}{x_0}
\end{equation}
for $x_0<0$.
From Eq.~(\ref{eq: coverage ordinary}), the convergence speed of the ordinary point series expansion is faster at larger $x_0$.
\subsection{Global Solution and Asymptotic Coefficients}\label{sec: Global Solution and Asymptotic Coefficients}
With the components introduced above, we can piece together the global solution of Eq.~(\ref{eq: P(x)}). Here is a brief description of this procedure.

\begin{enumerate}
\item Calculate asymptotic expansions at infinity and zero.
\item Obtain the values and derivatives at the edge of the valid range for these expansions.
\item Construct ordinary point expansions from the values and derivatives. This step extends the valid range of the asymptotic expansions at infinity or zero.
\item Perform the step 2 and step 3 until the valid range of asymptotic expansions at infinity and zero cover the chosen match point $x_m$ concurrently.
\item Calculate the connection coefficients of
\begin{align}
P^{\text{in}}_{0}&=C^{\text{in}}_{\infty}P^{\text{in}}_{\infty}+C^{\text{out}}_{\infty}P^{\text{out}}_{\infty},\label{eq: connect pin}\\
P^{\text{out}}_{\infty}&=C^{\text{in}}_{0}P^{\text{in}}_{0}+C^{\text{out}}_{0}P^{\text{out}}_{0}\label{eq: connect pout}
\end{align}
from the values and derivatives of each function at the match point $x_m$. Then, for $x<x_m$ and $x>x_m$, we can use Eq.~(\ref{eq: connect pin}) and Eq.~(\ref{eq: connect pout}) to give the results of $P^{\text{in}}_{0}$ and $P^{\text{out}}_{\infty}$, respectively.
\item Obtain $R^{\text{in}}$ and $R^{\text{up}}$ at arbitrary $x<0$ by
\begin{align}
R^{\text{in}}=TP^{\text{in}}_{0},\\
R^{\text{up}}=TP^{\text{out}}_{\infty}.
\end{align}
\end{enumerate}
\begin{figure}
\centering
\includegraphics[width=\linewidth]{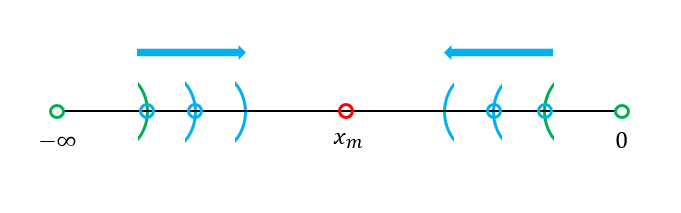}
\caption{A schematic diagram that describes the process of the global solution construction. The match point $x_m$ can be selected as an arbitrary number smaller than $0$.}
\label{fig: global_fig}
\end{figure}
See also Fig.~\ref{fig: global_fig}, which provides a schematic overview of the procedure above. In principle, the match point $x_m$ can be any point between $-\infty$ and $0$.

The simplest method to obtain the coefficients in Eq.~(\ref{eq: connect pin}) is solving the following equations at $x_m$
\begin{align}
P^{\text{in}}_{0}(x_m)&=C^{\text{in}}_{\infty}P^{\text{in}}_{\infty}(x_m)+C^{\text{out}}_{\infty}P^{\text{out}}_{\infty}(x_m),\\
{P^{\text{in}}_{0}}^\prime(x_m)&=C^{\text{in}}_{\infty}{P^{\text{in}}_{\infty}}^\prime(x_m)+C^{\text{out}}_{\infty}{P^{\text{out}}_{\infty}}^\prime(x_m)
\end{align}
and yield the result
\begin{align}
C^{\text{in}}_{\infty}&=\frac{\mathcal{I}(P^{\text{in}}_{0},P^{\text{out}}_{\infty},x_m)}{\mathcal{I}(P^{\text{in}}_{\infty},P^{\text{out}}_{\infty},x_m)},\label{eq: coe inner first}\\
C^{\text{out}}_{\infty}&=-\frac{\mathcal{I}(P^{\text{in}}_{0},P^{\text{in}}_{\infty},x_m)}{\mathcal{I}(P^{\text{in}}_{\infty},P^{\text{out}}_{\infty},x_m)},
\end{align}
where $\mathcal{I}$ is the inner product of the two functions at point $x$:
\begin{equation}
\mathcal{I}(P_1,P_2,x)=P_1(x)P_2^\prime(x)-P_1^\prime(x)P_2(x).
\end{equation}
A similar formula for coefficients in Eq.~(\ref{eq: connect pout}) can also be obtained as follows:
\begin{align}
C^{\text{in}}_{0}&=\frac{\mathcal{I}(P^{\text{out}}_{\infty},P^{\text{out}}_{0},x_m)}{\mathcal{I}(P^{\text{in}}_{0},P^{\text{out}}_{0},x_m)},\label{eq: coe inner final0}\\
C^{\text{out}}_{0}&=-\frac{\mathcal{I}(P^{\text{out}}_{\infty},P^{\text{in}}_{0},x_m)}{\mathcal{I}(P^{\text{in}}_{0},P^{\text{out}}_{0},x_m)}\label{eq: coe inner final}.
\end{align}

However, this intuitive approach will fail when the denominator is very close to $0$ (compare to the input number of the inner product), or in other words, the 2-dimensional vectors $(P^{\text{in}}_{\infty},{P^{\text{in}}_{\infty}}^\prime)$ and $(P^{\text{out}}_{\infty},{P^{\text{out}}_{\infty}}^\prime)$ nearly parallel to each other at $x_m$. Unfortunately, this is not a rare situation when we consider a large value of $\omega$. To overcome this question, we improve this method to make it work well in the nearly parallel case.

Note that for any two linear independent solutions for Eq.~(\ref{eq: P(x)}), $P_1(x)$ and $P_2(x)$, the following function is independent of $x$
\begin{equation}\label{eq: Wronskians}
\mathcal{W}_P(P_1,P_2)=\frac{\mathcal{I}(P_1,P_2,x)}{\exp{\left[-\int\frac{q_1(x)}{x(1-x)}\text{d}x\right]}}.
\end{equation}
We called $\mathcal{W}_P$ the Wronskian constant of Eq.~(\ref{eq: P(x)}). One can prove that the constant defined here is identical to the Wronskian constant of the Teukolsky radial equation $\mathcal{W}_T$. Here, the ``identity" means that the following condition holds:
\begin{equation}
\frac{\mathcal{W}_T(TP_1,TP_2)}{\mathcal{W}_P(P_1,P_2)}=\text{constant}
\end{equation}
for any two solutions of $P_1$, $P_2$. The constant can be scaled to $1$ by changing the integration constant in the denominator of Eq.~(\ref{eq: Wronskians}). In Appendix \ref{app: Identity of the Wronskians Constant}, we handle a more general case about the Wronskian constant under some transformation and include our situation as a special result.

In our situation, the Wronskian constant can be used to remove denominators of Eq.~(\ref{eq: coe inner final0}) and Eq.~(\ref{eq: coe inner final}). Note that
\begin{align}
\mathcal{W}_P(P^{\text{in}}_{0},P^{\text{out}}_{0})&=s+i(\epsilon+\tau),\\
\mathcal{W}_P(P^{\text{in}}_{\infty},P^{\text{out}}_{\infty})&=2i\epsilon\kappa.
\end{align}
These results are calculated at $x\rightarrow0$ and $x\rightarrow\infty$, respectively. Together with Eq.~(\ref{eq: coe inner first}-\ref{eq: coe inner final}), we obtain the following:
\begin{align}
&\iota=e^{2 i \kappa  x \epsilon } (1-x)^{s-i \tau +i \epsilon +1} (-x)^{ 1- s-i\tau -i\epsilon}\\
&C^{\text{in}}_{\infty}=\frac{\iota(x_m)}{2i\epsilon\kappa}\mathcal{I}(P^{\text{in}}_{0},P^{\text{out}}_{\infty},x_m),\label{eq: result coe p first}\\
&C^{\text{out}}_{\infty}=-\frac{\iota(x_m)}{2i\epsilon\kappa}\mathcal{I}(P^{\text{in}}_{0},P^{\text{in}}_{\infty},x_m),\\
&C^{\text{in}}_{0}=\frac{\iota(x_m)}{s+i(\epsilon+\tau)}\mathcal{I}(P^{\text{out}}_{\infty},P^{\text{out}}_{0},x_m),\\
&C^{\text{out}}_{0}=-\frac{\iota(x_m)}{s+i(\epsilon+\tau)}\mathcal{I}(P^{\text{out}}_{\infty},P^{\text{in}}_{0},x_m),\label{eq: result coe p last}
\end{align}
The choice of this integrated constant or $x_m$ can not have any influence on our results Eq.~(\ref{eq: result coe p first})-(\ref{eq: result coe p last}).

Since the asymptotic coefficients in Eq.~(\ref{eq: rin bound condition}) ans Eq.~(\ref{eq: rup bound condition}) are necessary for constructing the Green function of the Teukolsky radial equation, we give the relation between the coefficients in the above equations and the ones in Eq.~(\ref{eq: connect pin}-\ref{eq: connect pout}) by considering the limit of certain fractions at $0$ or $\infty$,
\begin{align}
\mathbf{f}_1&=\lim_{r\rightarrow r_+}\frac{\Delta^{-s}e^{-ipr_*}}{TP^{\text{in}}_0}=4^s(1-a^2)^{ip+s}e^{ipr_+},\\
\mathbf{f}_2&=\lim_{r\rightarrow r_+}\frac{e^{-ipr_*}}{T P^{\text{out}}_0}=(1-a^2)^{-ip}e^{-ipr_+},\\
\mathbf{f}_3&=\lim_{r\rightarrow\infty}\frac{e^{-i\omega r_*}}{rTP^{\text{in}}_\infty}=2(1-a^2)^{i\omega+1/2}e^{i\omega r_+},\\
\mathbf{f}_4&=\lim_{r\rightarrow\infty}\frac{e^{i\omega r_*}}{r^{2s+1}TP^{\text{out}}_\infty}=2(1-a^2)^{s-i\omega+1/2}e^{-i\omega r_+}.
\end{align}
Thus, we have
\begin{equation}\label{eq: asy coe se}
\begin{aligned}
B^{\text{trans}}&=\mathbf{f}_1,&C^{\text{trans}}&=\mathbf{f}_4,\\
B^{\text{inc}}&=C^{\text{in}}_\infty\mathbf{f}_3,&C^{\text{ref}}&=C^{\text{in}}_0\mathbf{f}_1,\\
B^{\text{ref}}&=C^{\text{out}}_\infty\mathbf{f}_4,&C^{\text{inc}}&=C^{\text{out}}_0\mathbf{f}_2.
\end{aligned}
\end{equation}
With the asymptotic series introduced in Sec.~\ref{sec: Horizon Solution} and Sec.~\ref{sec: Infinity Solution}, as well as the ordinary point power series expansion in Sec.~\ref{sec: Ordinary  Point Expansion}, we construct the global solution and obtain the asymptotic coefficients. However, the performance of the infinity asymptotic series are pretty poor when $\omega$ very close to $0$ ($|\omega|<0.01$) and in which the value of $\omega$ the confluent hypergeometric function expansion should be an idea candidate for the calculation of the homogeneous solution, see Appendix \ref{app: Confluent Hypergeometric Expansion} for detail. Meanwhile, the ordinary point expansion and the Frobenius solution near the horizon will lose both efficiency and accuracy when encountering a large $\omega$. Thus, in the next section, we solve this problem with the large frequency expansion.

\section{Large Frequency Expansion}\label{sec: Large Frequency Expansion}
In the previous sections, we introduced a new method for calculating the homogeneous solution. However, they are not suitable for large $\omega$, the high-frequency oscillation will make the ordinary point expansion go back to the numerical integration, and the convergence speed of the asymptotic series around the horizon will become unacceptably slow. These facts motivate us to find a more efficient method for large $\omega$.

We begin with Eq.~(\ref{eq: minor changed teukolsk equation}) and also set $x=(r_+-r)/(r_+-r_-)$. In contrast to the previous section, the transformation $R(r)=E(x)X(x)$ is used to eliminate the coefficient of $R^\prime(r)$ in Eq.~(\ref{eq: minor changed teukolsk equation}), the result reads
\begin{align}
X^{\prime\prime}=&\beta^2 \tilde{f}X,\label{eq: first changed teukolsky equation}\\
\tilde{f}(x)=&f(x)/\beta^2\\
f(x)=&\frac{\alpha_0+\alpha_1 x+\alpha_2 x^2+\alpha_3 x^3+\alpha_4 x^4}{(1-x)^2x^2},\label{eq: large omega f(x)}\\
E(x)=&(1-x)^{-(s+1)/2}(-x)^{-(s+1)/2},
\end{align}
with
\begin{align}
\alpha_0=&\frac{\kappa ^2 \left(s^2-1\right)-2 i \kappa  s \ttau -\ttau ^2}{4 \kappa ^2}\nonumber\\
&+\frac{i (\kappa +1)  (\kappa  s-i \ttau ) \omega}{\kappa ^2}-\frac{(\kappa +1)^2 \omega ^2}{\kappa ^2}\\
\alpha_1=&\left(\frac{4 (\kappa +1)^2-\tlam}{\kappa }\right)\omega ^2 -\frac{2 (\kappa +1)  (\ttau -i (\kappa -1) s) \omega}{\kappa }\nonumber\\
&+s \left(\frac{i \ttau }{\kappa }-s-1\right)\\
\alpha_2=&\omega ^2 (\tlam - 4 \kappa  (\kappa +3)-8)+\omega  (2 \ttau -6 i \kappa  s)+s (s+1)\\
\alpha_3=&8 \kappa  (\kappa +1) \omega ^2+4 i \kappa  s \omega\\
\alpha_4=&-4 \kappa ^2 \omega ^2
\end{align}
where $\kappa=\sqrt{1-a^2}$, $\ttau=ma$, the complex number $\beta$ with a positive real part indicates the scale of $f$, and $\tilde{\lambda}=\omega^{-2}\lambda$ is introduced due to $\lambda\sim\mathcal{O}(\omega^2)$ when $\omega$ is real. The details can be found in Ref.~\cite{PhysRevD.73.024013}.

For $x_0\neq 0,1$ and $f(x_0)\neq 0$, to give an explicit form of the local solution around $x_0$ for Eq.~(\ref{eq: first changed teukolsky equation}), we transform $X$ to $\tilde{f}^{-1/4}\chi$ and perform the following change of $x$
\begin{equation}\label{eq: eta in x}
\eta(x)=\int \sqrt{\tf(x)}dx
\end{equation}
then we can bring Eq.~(\ref{eq: first changed teukolsky equation}) into a ``standard" form
\begin{equation}\label{eq: pure wave function1}
\chi^{\prime\prime}(\eta)=[\beta^2+\phi(\eta)]\chi(\eta),
\end{equation}
where
\begin{equation}
\phi(\eta(x))=\frac{4\tf(x)\tf^{\prime\prime}(x)-5\tf^\prime(x)^2}{16\tf(x)^3}
\end{equation}
The two solution of the Eq.~(\ref{eq: pure wave function1}) are \cite{OLVER1974362}
\begin{align}
\chi_1(x)=&e^{\beta\eta(x)}\sum_{n=0}^\infty\frac{A_n(x)}{\beta^n}\\
\chi_2(x)=&e^{-\beta\eta(x)}\sum_{n=0}^\infty(-1)^n\frac{A_n(x)}{\beta^n}
\end{align}
with $A_0=1$ and
\begin{equation}\label{eq: integrate expression}
A_{s+1}(x)=-\frac{1}{2}\frac{A_s^\prime(x)}{\tf(x)^{1/2}}+\frac{1}{2}\int\phi(\eta(x))\tf(x)^{1/2}A_s(x)dx
\end{equation}
It is quite difficult to find (may not exist) an analytic form of the integral~(\ref{eq: integrate expression}). Therefore, similar to the technique developed in Sec.~\ref{sec: Global Solution and Asymptotic Coefficients}, the series expansion is given instead. 

For convenience, we set $\beta^2=f(x_0)$ and $\tf(x)=1+\sum_{i=1}^\infty f_n \tilde{x}^n$
with $\tx=x-x_0$. Here, terms up to $\mathcal{O}(\tx^4)$ of the ordinary point expansion are
\begin{align}
A_1=&\frac{\left(-175 f_1^4+620 f_2 f_1^2-528 f_3 f_1-256 f_2^2+384 f_4\right)}{768}  \tx^3\nonumber\\
&+\frac{1}{128} \left(25 f_1^3-64 f_2 f_1+48 f_3\right) \tx^2\nonumber\\
&+\frac{1}{32} \left(8 f_2-5 f_1^2\right) \tx,\\
A_2=&\left(\frac{985 f_1^4}{2048}-\frac{203 f_1^2 f_2}{128} +\frac{39 f_1 f_3+19 f_2^2-24 f_4}{32}\right)\tx^2\nonumber\\
&-\frac{3}{64} \left(5 f_1^3-12 f_2 f_1+8 f_3\right) \tx+\frac{1}{64} \left(5 f_1^2-8 f_2\right),\\
A_3=&\frac{-565 f_1^4+1808 f_2 f_1^2-1344 f_3 f_1-640 f_2^2+768 f_4}{1024} \tx\nonumber\\
&+\frac{3}{128} \left(5 f_1^3-12 f_2 f_1+8 f_3\right),\\
A_4=&\frac{565 f_1^4-1808 f_2 f_1^2+1344 f_3 f_1+640 f_2^2-768 f_4}{2048}.
\end{align}
The derivative in Eq.~(\ref{eq: integrate expression}) connects the coefficient of $x^m$ in $A_n$ with the $x^{m-1}$ in $A_{n+1}$. Consequentially, if $A_1$ is calculated to $m$-th term, then $A_n$ can only accuracy to $(m-n+1)$-th term. However, in practice, we can restrict the valid range of the above expansion to be smaller than $|\beta|$ and then can only reserve the $(m-n+1)$-th term for $A_n$ safely.

For the local solution at $x=0$, following Ref.~\cite{ferreira2014olversasymptoticmethodspecial}, we transform $X$ to $(d\xi/dx)^{-1/2}\zeta$ and perform the following change of $x$
\begin{equation}
\xi(x)=\exp{\int \sqrt{\tf(x)} dx}
\end{equation}
then we can bring Eq.~(\ref{eq: first changed teukolsky equation}) into a ``standard" form,
\begin{equation}\label{eq: pure wave function}
\zeta^{\prime\prime}(\xi)=\left[\frac{\beta^2}{\xi^2}+\psi(\xi)\right]\zeta(\xi),
\end{equation}
where
\begin{equation}
\begin{aligned}
\psi(\xi(x))=&\exp{\left[-2 \int \sqrt{\tf(x)} \, dx\right]}\times\\
&\frac{ 4 \tf(x) \tf''(x)-5 \tf'(x)^2-4 \tf(x)^3}{16 \tf(x)^3}.
\end{aligned}
\end{equation}
The two solutions of the differential equation above are
\begin{align}
\zeta_1(x)&=\xi(x)^\Lambda\sum_{n=0}^\infty\frac{B_n(x)}{2^n\Lambda^n}\\
\zeta_2(x)&=\xi(x)^{1-\Lambda}\sum_{n=0}^\infty\frac{B_n(x)}{2^n(1-\Lambda)^n}
\end{align}
with $\Lambda(\Lambda-1)=\beta^2$, $B_0=1$, and
\begin{equation}
\begin{aligned}
B_{n+1}(x)=&B_n(x)-B_n^\prime(x)\frac{\xi(x)}{\xi(x)^\prime}\\
&+\int B_n(x)\psi(\xi(x))\xi(x)\xi(x)^\prime\,dx
\end{aligned}
\end{equation}
Similar to the discussion above, we also give a power series expression. For convenience, we set $\beta^2=\alpha_0$ and $\gamma_i=\alpha_i/\alpha_0$. Therefore, terms up to $\mathcal{O}(\tx^4)$ are
\begin{widetext}
\begin{align}
A_1(x)=&\frac{1}{96} \left(-37 \gamma _1^3-38 \gamma _1^2+\left(92 \gamma _2+8\right) \gamma _1+8 \left(7 \gamma _2-8 \gamma _3\right)\right) x^3+\frac{3}{32} \left(3 \gamma _1^2+4 \gamma _1-4 \gamma _2\right) x^2+1\\
A_2(x)=&\frac{1}{24} \left(-53 \gamma _1^3-70 \gamma _1^2+4 \left(31 \gamma _2-2\right) \gamma _1+88 \gamma _2-80 \gamma _3\right) x^3+\frac{3}{8} \left(3 \gamma _1^2+4 \gamma _1-4 \gamma _2\right) x^2+1\\
A_3(x)=&\frac{1}{32} \left(-331 \gamma _1^3-506 \gamma _1^2+4 \left(185 \gamma _2-34\right) \gamma _1+8 \left(73 \gamma _2-56 \gamma _3\right)\right) x^3+\frac{39}{32} \left(3 \gamma _1^2+4 \gamma _1-4 \gamma _2\right) x^2+1\\
A_4(x)=&-\frac{5}{12} \left(109 \gamma _1^3+182 \gamma _1^2+\left(64-236 \gamma _2\right) \gamma _1-200 \gamma _2+136 \gamma _3\right) x^3+\frac{15}{4} \left(3 \gamma _1^2+4 \gamma _1-4 \gamma _2\right) x^2+1
\end{align}
\end{widetext}

As discussed in the previous section, the calculation of the asymptotic coefficients depends on the asymptotic behavior of $\zeta_{1,2}(x)$ when $x\rightarrow0$. We have
\begin{align}
&\lim_{x\rightarrow 0}\frac{E(x)(d\xi/dx)^{-1/2}\zeta_1(x)}{T(x)P_{0}^{\text{in}}(x)}\nonumber\\
&=-i^{-s}\exp{\frac{\pi(\ttau-2(1+\kappa)\omega)}{2\kappa}}\sum_{n=0}^\infty\frac{B_n(0)}{2^n\Lambda^n},\\
&\lim_{x\rightarrow 0}\frac{E(x)(d\xi/dx)^{-1/2}\zeta_2(x)}{T(x)P_{0}^{\text{out}}(x)}\nonumber\\
&=-\exp{\frac{\pi(2(1+\kappa)\omega-\ttau-3\kappa s)}{2\kappa}}\sum_{n=0}^\infty\frac{B_n(0)}{2^n(1-\Lambda)^n}.
\end{align}
Note that $B_n(0)=1$ for arbitrary $n$, we have
\begin{align}
&\sum_{n=0}^\infty\frac{B_n(0)}{2^n\Lambda^n}=\frac{2\Lambda}{2\Lambda-1}\\
&\sum_{n=0}^\infty\frac{B_n(0)}{2^n(1-\Lambda)^n}=\frac{2\Lambda-2}{2\Lambda-1}
\end{align}
Moreover, the convergence range of the large frequency expansion is restricted by the zeros and singularities of $f(x)$, see the discussion in Appendix \ref{app: Transition Points}.
\section{Numerical Experiment}\label{sec: Numerical Experiment}
The numerical waveforms for EMRIs require a huge amount of homogeneous solutions of the Teukolsky equation. Thus, both accuracy and efficiency are critical for calculating the homogeneous solutions. In this section, we perform a series of numerical experiments and compare our method with the MST and SN methods.  Our implementation is written in \texttt{rust}, and tests on the Intel Xeon Gold 6342 CPU 2.80GHz platform with a single thread.

We first demonstrate the accuracy that our method can achieve (with 64-bit floating-point numbers). If a function $R(r)$ is the solution of Eq.~(\ref{eq: minor changed teukolsk equation}), after putting $R(r)$ back into Eq.~(\ref{eq: minor changed teukolsk equation}), the result will be zero. Thus, the following residual $\varepsilon(R,r)$ is defined to indicate the relative error of solutions $R$ at $r$:
\begin{equation}\label{eq: residual equation}
\begin{aligned}
&\varepsilon(R,r)\\
&=\frac{\left|\Delta{R}''(r)+2(s+1)(r-1)R'(r)-V(r)R(r)\right|}{\max\{|\Delta{R}''(r)|,|2(s+1)(r-1)R'(r)|,|V(r)R(r)|\}}
\end{aligned}
\end{equation}
\begin{figure}
\centering
\includegraphics[width=\linewidth]{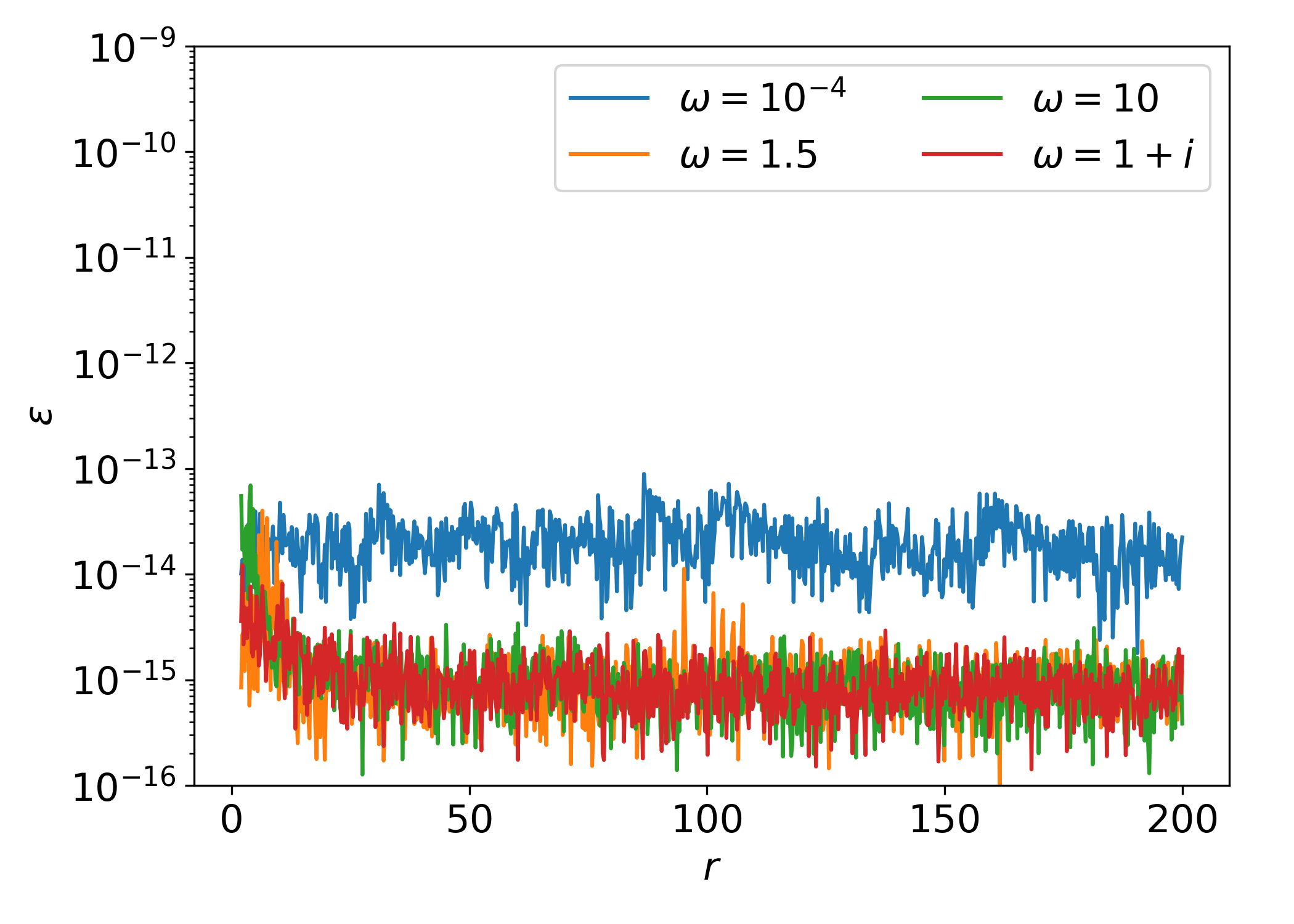}
\caption{Residual $\varepsilon=\max\{\varepsilon(R^{\text{in}}),\varepsilon(R^{\text{up}})\}$ for $s=-2,l=m=2,a=0.99$ from $r=2$ to $200$. The relative errors are smaller than $10^{-13}$.} 
\label{fig: error}
\end{figure}
Fig.~\ref{fig: error} shows the results of $\varepsilon$ with $s=-2$, $l=m=2$, $a=0.99$. The relative errors are approximately smaller than $10^{-13}$, no matter whether $\omega$ is small or large, real or complex. The trend for $\omega=10^{-4}$ is different from the others because the confluent hypergeometric expansion is used when $|\omega|<10^{-2}$. Although our method performed well in this check, we still need an additional test. Because $\varepsilon$ can only represent the local property of solutions, e.g., assuming a new solution $R^{\text{sp}}(r)$ constructed by splicing a fragment of $R^{\text{in}}$ into $R^{\text{up}}$, then $\varepsilon(R^{\text{sp}},r)$ definitely is a small value, but $R^{\text{sp}}(r)$ is not the solution we need.

To resolve this problem, we introduce a consistency check by using the Wronskian constant. The Wronskian constant should be the same everywhere; thus, we define the deviation $\varpi$ as
\begin{equation}
\varpi(r)=\left|\frac{\mathcal{W}(r)-2i\omega B^{\text{inc}}C^{\text{trans}}}{\mathcal{W}(r)}\right|
\end{equation}
where $\mathcal{W}(r)=\mathcal{I}(R^{\text{in}},R^{\text{up}},r)/\Delta^{s+1}$ and $2i\omega B^{\text{inc}}C^{\text{trans}}$ is the Wronskian constant calculated analytically at $r=\infty$. If the solutions we obtained exactly are $R^{\text{in}}$, $R^{\text{up}}$ then $\varpi$ should equal to zero.
\begin{figure}
\centering
\includegraphics[width=\linewidth]{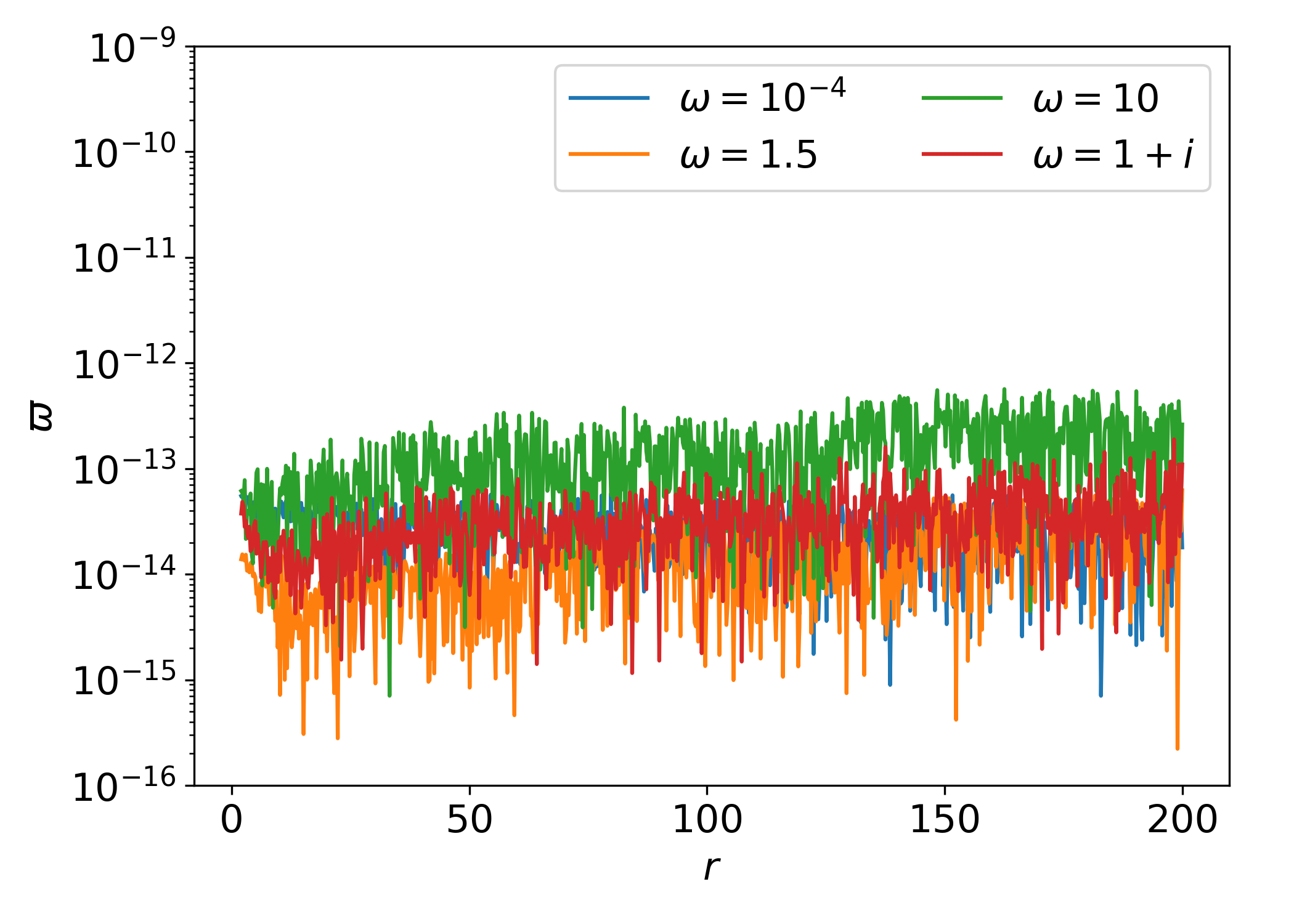}
\caption{The deviation $\varpi$ with $s=-2$, $l=m=2$, $a=0.99$ from $r=2$ to $200$. The deviations stay roughly at $\varpi\approx10^{-13}$}
\label{fig: self_consistant_check}
\end{figure}
Fig.~\ref{fig: self_consistant_check} shows the deviation $\varpi$ with the same parameters set in Fig.~\ref{fig: error}. The deviations are smaller than $10^{-12}$, no matter whether $\omega$ is small or large, real or complex. Additionally, Fig.~\ref{fig: self_consistant_check} also confirms the correctness of the asymptotic coefficients we obtained.

Now, we show the efficiency. We perform a rough test to illustrate the computation time of our JH method with varied parameters. The test calculates the values of $R^{\text{in}}$, $R^{\text{up}}$ and their derivatives at the 1000 points from $2r_+$ to $20r_+$. This test will be repeated 10 times to get the average computation time. 
\begin{figure}
\centering
\includegraphics[width=\linewidth]{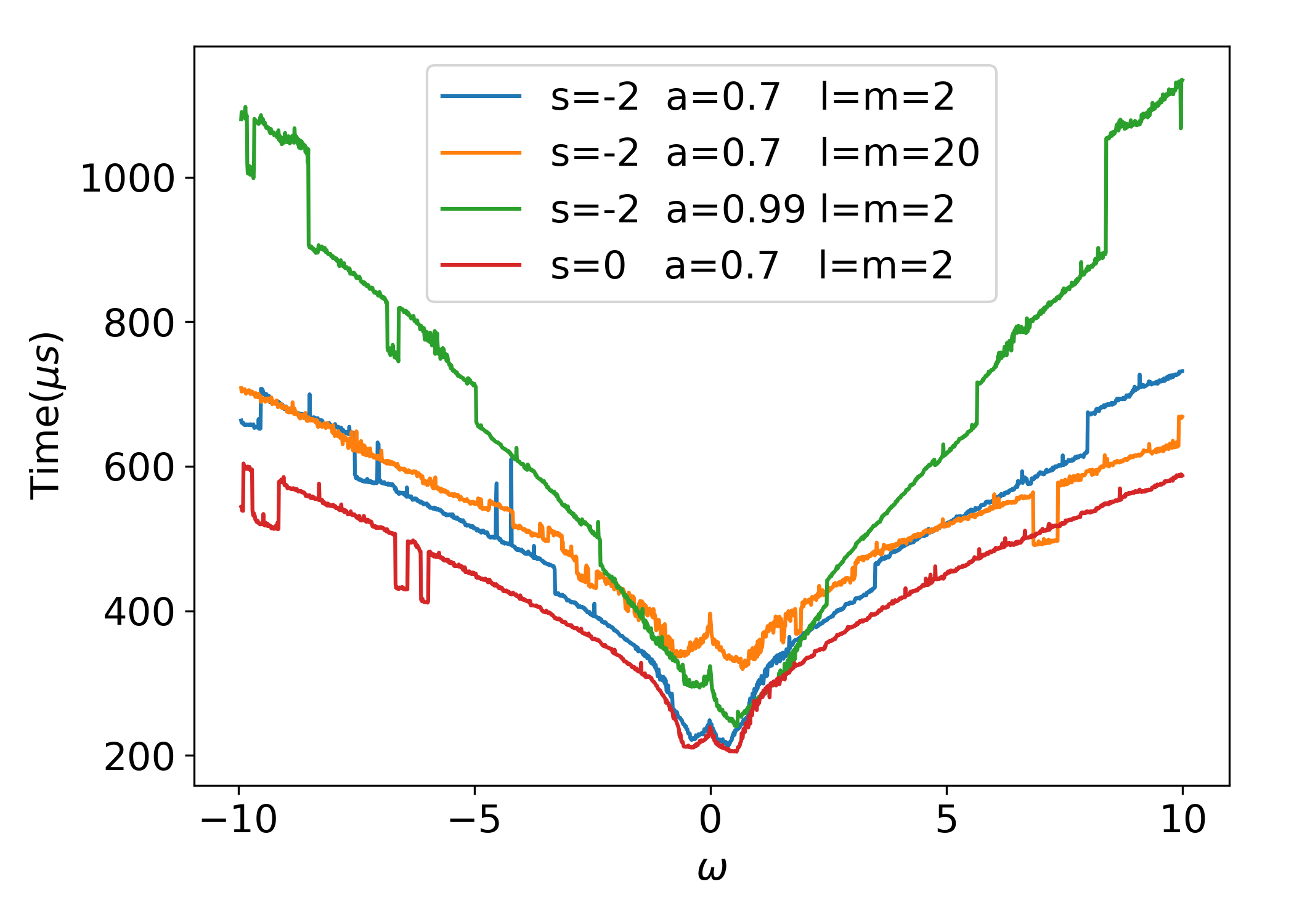}
\caption{The computation time of 1000 set of $R^{\text{in}}$, $R^{\text{up}}$ and their derivatives on $2r_+$ to $20r_+$.} 
\label{fig: SE_time_average}
\end{figure}

From Fig.~\ref{fig: SE_time_average}, we can find that the computation time is between 0.2 ms and 1 ms even for extreme spin and large $|\omega|$. When $a$ is close to $1$, the two singular points tend to merge, which changes the property of the solution near $x=0$, make the computation time a little longer. However, the efficiency of the JH method is much higher than the MST and SN ones, we will show the details as follows. 


As mentioned in Sec.~\ref{sec: Introduction}, the homogeneous Teukolsky equation can be solved in several ways, e.g., the MST and SN methods. A more detailed test will be conducted to compare the efficiency of the different methods. For the MST method, we choose the \texttt{Gremlin} in Black Hole Perturbation Toolkit \cite{BHPToolkit}. \texttt{Gremlin} is a toolkit for studying solutions to the Teukolsky equation using a point-particle source, which is implemented in \texttt{C++}. For the SN method, we choose the \texttt{GeneralizedSasakiNakamura.jl} described in Ref.~\cite{PhysRevD.110.124070,Lo:2025njp}. It is a high-performance package written in \texttt{julia}.

The computation expense can be divided into two parts. The first part is the initialization time $t_i$. In the actual implementation, we usually construct a structure to store the intermediate data. When calculating at some $r$, we can use this structure to avoid repeat computations. The time required to generate this structure is referred to as the initialization time. Usually, asymptotic coefficients are also obtained in this stage. The second part is the time cost on the calculation for the value at a single point $r$, denoted by $t_c$. 

\begin{table*}
\begin{tabularx}{\textwidth}{XXXXXXXXXX}
\toprule[1.5pt]
& \multicolumn{3}{c}{MST} & \multicolumn{3}{c}{SN} & \multicolumn{3}{c}{JH} \\ \midrule
$\omega$ &    $10^{-4}$    &    $10^{-2}$    &   $1$    &    $10^{-4}$    &   $10^{-2}$    &    $1$   &    $10^{-4}$    &    $10^{-2}$   &   $1$    \\ \midrule
$t_i$    &    16.2 ms    &    24.5 ms    &   52.4 ms    &    22.4 ms    &   46.4 ms    &   165 ms    &    32.8 $\mu$s    &   37.5 $\mu$s    &   49.5 $\mu$s    \\ \midrule
$\bar{t}_c$    &    3.66 ms    &    5.53 ms    &   97.1 ms    &    12.3 $\mu$s    &   11.9 $\mu$s    &   11.5 $\mu$s    &    108 ns    &    107 ns   &    136 ns   \\ \bottomrule[1.5pt]
\end{tabularx}
\caption{The comparison of computation efficiency for the three methods, $t_i$ is the initialization time, and $\bar{t}_c$ is the average calculation time for a single point. We fix $s=-2$, $a=0.7$, and $l=m=2$.}
\label{tab: time table}
\end{table*}
Table.~\ref{tab: time table} clearly shows the advantages of JH method, about three order faster than the other two methods. Our method is better than the MST method in initialization time $t_i$ because the MST method involves arbitrary precision arithmetic. Compared to the SN method, due to the absence of the SN transformation in our method, we achieved a 100-fold reduction in the average calculation time $\bar{t}_c$. Moreover, the MST method performs poorly in terms of the average calculation time $\bar{t}_c$, in fact, this is in line with expectations. Since the calculation of the hypergeometric functions is much more expensive than interpolating the numerically integrated data.

\section{Conclusion}\label{sec: Conclusion}
In this paper, we propose a new numerical method by integrating local analytic expansions at various points and asymptotic expansions at singular points. We have also disposed of cases with a small or large frequency. Overall, our method significantly enhances computational efficiency while maintaining accuracy. Compared with the existing methods (MST, SN), our JH approach can cover a wider frequency range (even for complex frequencies) and achieve a huge improvement in computational efficiency.

An important application of our new method is Teukolsky-based waveform generation. The numerical calculation of the EMRIs waveform performs a Fourier expansion for the source. The frequency is composed by the radial, polar and azimuthal parts. For a generic orbit, each ingredient of the waveform must have a frequency $\Omega=m\Omega_\phi+k\Omega_\theta+n\Omega_n$ for certain harmonic numbers $m,k,n$ \cite{PhysRevD.73.024027}. Moreover, the index $l$ generated by the parameter separation should also be considered.

With these four discrete dimensions, in order to achieve the accuracy requirement, the number of harmonics needed to be calculated can easily reach tens of thousands or even millions for high eccentricity and high inclination orbits. Hence, homogeneous solution generation is the key performance bottleneck in the entire process. Using our new method, the calculation of Teukolsky-based waveforms will have a significant increase in efficiency. 

Another potential application is the Kerr black hole quasi-normal mode with a modified boundary condition. By posing the condition that $C^{\text{inc}}$ or $B^{\text{inc}}$ is equal to zero, the quasi-normal frequency can be viewed as a root finding question. Analogously, one can also obtain the quasi-normal mode under any modified boundary condition at the horizon and infinity. However, for those who are only concerned with the standard boundary condition, we still suggest Leaver's method \cite{doi:10.1098/rspa.1985.0119, 10.1063/1.527130}, which directly obtains the quasi-normal mode by solving a continued fraction equation.

In conclusion, due to the well performance of the JH method, it will be very useful for the waveform template construction of EMRIs and ringdown analysis for exotic compact objects.
\begin{acknowledgements}
This work is supported by The National Key R\&D Program of China(GrantNo. 2021YFC2203002),
NSFC (National Natural Science Foundation of China) No. 12473075, No. 12173071 and No.
11773059. This work made use of the High Performance Computing Resource in the Core
Facility for Advanced Research Computing at Shanghai Astronomical Observatory. This work makes use of the Black Hole Perturbation Toolkit.
\end{acknowledgements}
\appendix
\section{Angular Teukolsky Equation}\label{app: Angular Teukolsky Equation}
The Teukolsky equation can be separated into two parts; the radial part has been discussed in the main text, and this appendix focuses on the angular part.

By setting $x=\cos\theta$ and $c=a\omega$, the angular Teukolsky equation becomes the following:
\begin{equation}
\begin{aligned}
&\frac{\text{d}}{\text{d}x}\left[(1-x^2)\frac{\text{d}}{\text{d}x}S(x)\right]+V_\theta(x) S(x)=0\\
&V_\theta(x)=(cx)^2-2csx+s+\mathcal{A}-\frac{(m+sx)^2}{1-x^2}
\end{aligned}
\end{equation}
where $\mathcal{A}$ is relate to $\lambda$ in Eq.~(\ref{eq: teukolsky potential}) by
\begin{equation}
\lambda=\mathcal{A}+c^2-2mc
\end{equation}
The solutions we found should be finite at $x=\pm1$. With these boundary conditions, the solutions are the spin-weighted spheroidal harmonics ${_sS_{lm}}(x;c)$. There exist a variety of methods to obtain the spin-weighted spheroidal harmonics numerically, such as Leaver's continued fraction method \cite{doi:10.1098/rspa.1985.0119} and the spectral decomposition method \cite{PhysRevD.61.084004, PhysRevD.90.124021}. The continued fraction method only determines the eigenfunction up to a normalization constant, which must be fixed by imposing the normalization condition
\begin{equation}
\int^{1}_{-1}|{_sS_{lm}}(x;c)|^2\text{d}x=1
\end{equation}
numerically. This step is too expensive compared with the spectral decomposition method, especially for the high-frequency oscillation mode. The spectral decomposition method performs the following expansion:
\begin{equation}\label{eq: spectral decomposition}
{_sS_{lm}}(x;c)=\sum_{l^\prime=l_{\text{min}}}^\infty{_sC_{l^\prime m}}(c){_sY_{lm}(x)}
\end{equation}
where $l_{\text{min}}=\max{(|m|,|s|)}$, $_sY_{lm}(x)={_sS_{lm}}(x;0)$. The calculation of the coefficients can be found in Ref.~\cite{PhysRevD.90.124021}, we do not repeat it here. This formula can be easily normalized by the orthogonal condition of $_sY_{lm}(x)$
\begin{equation}
\int^{1}_{-1}{_sY_{lm}(x)}{_sY_{l^\prime m}(x)}^*\text{d}x=\delta_{ll^\prime},
\end{equation}
which is equal to require
\begin{equation}
\sum_{l^\prime=l_{\text{min}}}^\infty|_sC_{l^\prime m}|^2=1.
\end{equation}
We chose to adopt the spectral decomposition method in practice because of the discussion above. Thus, we need the expansion coefficients ${_sC_{l^\prime m}}(c)$ and the spin-weighted spherical harmonics ${_sY_{lm}(x)}$ in Eq.~(\ref{eq: spectral decomposition}) to construct the solution ${_sS_{lm}}(x;c)$. The expansion coefficients ${_sC_{l^\prime m}}(c)$ have been well discussed in a series of studies. Here, we focus on the numerical calculation of the spin-weighted spherical harmonics. The close form of ${_sY_{lm}(x)}$ can be written as follows:
\begin{equation}\label{eq: close form Y}
_sY_{lm}(x)=A_{n}^{(k_+,k_-)}(1-x)^{k_+/2}(1+x)^{k_-/2}J_n^{(k_+,k_-)}(x)
\end{equation}
where $k_+=|m+s|$, $k_-=|m-s|$, $n=l-l_{\text{min}}$, and the prefix normalization factor
\begin{equation}
\begin{aligned}
A_{n}^{(k_+,k_-)}=&\sqrt{\frac{2n+k_++k_-+1}{2^{k_++k_-+1}}}\\
&\times\sqrt{\frac{\Gamma(n+k_++k_-+1)n!}{\Gamma(n+k_++1)\Gamma(n+k_-+1)}}.\label{eq: sph prefix}
\end{aligned}
\end{equation}
The function $J_n^{(k_+,k_-)}(x)$ introduced here is the Jacobi polynomial
\begin{equation}\label{eq: close form J}
\begin{aligned}
&J_n^{(k_+,k_-)}(x)\\
=&\sum_{r=0}^n\left( \begin{array}{c} n+k_+ \\ n-r \end{array} \right)\left( \begin{array}{c} n+k_- \\ r \end{array} \right)\left( \frac{x-1}{2} \right)^r\left( \frac{x+1}{2} \right)^{n-r},
\end{aligned}
\end{equation}
However, it is not a good choice to directly evaluate the closed form of $_sY_{lm}$ by Eq.~(\ref{eq: close form Y})-(\ref{eq: close form J}). Not only because of the $\mathcal{O}(n^2)$ time complexity but also because the large factorials involved in the computation may trigger an overflow error. Thus, we suggest adopting the following recurrence relation of $J_n^{(k_+,k_-)}(x)$ in the calculation of the value and derivative of ${_sS_{lm}}(x;c)$:
\begin{align}
&J_0^{(k_+,k_-)}(x)=1\\
&J_1^{(k_+,k_-)}(x)=(k_++1)+(k_++k_-+2)\frac{x-1}{2}\\
&J_n^{(k_+,k_-)}(x)\nonumber\\
&=\frac{(c-1)(c(c-2)x+(a-b)(c-2n))}{2n(c-n)(c-2)}J_{n-1}^{(k_+,k_-)}(x)\\
&\quad-\frac{(a-1)(b-1)}{n(c-n)(c-2)}cJ_{n-2}^{(k_+,k_-)}(x)\nonumber\\
&\frac{\text{d}J_n^{(k_+,k_-)}(x)}{\text{d}x}\nonumber\\
&=\frac{b-a+cx}{c(x^2-1)}J_n^{(k_+,k_-)}(x)-\frac{2ab}{c(x^2-1)}J_{n-1}^{(k_+,k_-)}(x)\label{eq: sph derivative}
\end{align}
where $a=k_++n$, $b=k_-+n$ and $c=k_++k_-+2n$.
\section{Confluent Hypergeometric Expansion}\label{app: Confluent Hypergeometric Expansion}
Since the type of the singular point changes when $\omega$ is close to $0$, the asymptotic series expansion will be troublesome in this case. This appendix introduces another expansion to solve this problem. For the two local solutions at $\infty$ of Eq.~(\ref{eq: P(x)}) appeared as an asymptotic series before, we present the following form in the confluent hypergeometric function as follows
\begin{equation}
\begin{aligned}
P^{\text{in}}_\infty=&\mathbb{F}^{\text{in}}_\infty(1-x)^{-i\epsilon+s+i\tau-1}\sum_{n=-\infty}^{+\infty} f^\nu_nz^{n+\nu+1-s+i\epsilon}\\&\times U(n+\nu+1-s+i\epsilon,2n+2\nu+2;z),\\
P^{\text{out}}_\infty=&\mathbb{F}^{\text{out}}_\infty e^{-2i\epsilon\kappa x}(1-x)^{i\epsilon-s+i\tau-1}\sum_{n=-\infty}^{+\infty} f^\nu_nz^{n+\nu+1+s-i\epsilon} \\&\times\frac{(\nu+1+s-i\epsilon)_n}{(\nu+1-s+i\epsilon)_n}\\&\times U(n+\nu+1+s-i\epsilon,2n+2\nu+2;-z)\,,\label{eq: confluent hyper expansion}
\end{aligned}
\end{equation}
where $z=2i\epsilon\kappa(1-x)$, $f_n^\nu$ satisfy a three-term recurrence relation
\begin{equation}\label{eq: three-term recurrence relation}
\alpha^\nu_na_{n+1}+\beta^\nu_na_n+\gamma_n^\nu a_{n-1}=0,
\end{equation}
$U$ is the irregular confluent hypergeometric function, $\alpha^\nu_n$, $\beta^\nu_n$ and $\gamma^\nu_n$ are the same as the corresponding values in the MST method, see Ref.~\cite{10.1143/PTP.112.415}. The prefix in Eq.~(\ref{eq: confluent hyper expansion})
\begin{align}
\mathbb{F}^{\text{in}}_\infty&=\sum_{n=-\infty}^{+\infty} f^\nu_n\\
\mathbb{F}^{\text{out}}_\infty&=\sum_{n=-\infty}^{+\infty} f^\nu_ne^{i\pi(n+\nu+1+s-i\epsilon)}\frac{(\nu+1+s-i\epsilon)_n}{(\nu+1-s+i\epsilon)_n}
\end{align}
is set to scale the new formula and make it agree with the asymptotic series in the main text.
Here, parameter $\nu$ is introduced to ensure that $f_n^\nu$ is the minimal solution of the recurrence relation in both directions; consequently, $f_n^\nu$ will converge when $n\rightarrow-\infty$ and $n\rightarrow+\infty$ at the same time.

For calculating the minimal solution, we suggest Olver's method \cite{olwr71numerical}, which achieves the $\mathcal{O}(n)$ time complexity for the error estimate.
\section{Derivatives of the Homogeneous Solutions of the Radial and Angular Teukolsky Equation}
The derivatives of the homogeneous solutions are needed to evaluate the source terms $\mathcal{S}(T^{\mu\nu})$. In this appendix, we introduce a recursive method to obtain arbitrary high-order derivatives of the homogeneous solutions $R_{lm\omega}(r)$.
We first discuss the first-order derivative. For radial homogeneous solutions $R_{lm\omega}(r)$, because of the polynomial expression, things are quite easy; in the MST method, we adopt the following equation for the calculation of derivatives:
\begin{align}
&\frac{\text{d}{U(a,b;x)}}{\text{d}x}\nonumber\\
&=\frac{(1+a-b)}{z(b-2)}U(a-1,b-2;x)\\
&\quad-\frac{(b-1)(b-2)+z(1+a-b)}{z(b-2)}U(a,b;x),\nonumber
\end{align}
where $(a,b,c,x)$ is an arbitrary couple of parameters. With these relationships, it is sufficient to obtain the derivatives of $R^{\text{in}}$ and $R^{\text{up}}$.

One can also obtain the higher-order derivatives of $R(r)$ by adopting the ordinary point expansion described in the main text; however, the prefix $T(x)$ is quite awful, because this term makes the time complex of the calculation become $\mathcal{O}(n^2)$, where $n$ is the order of the derivative to be calculated. Here, with the following definition for any smooth function $f$
\begin{equation}\label{eq: derivative notation}
\begin{aligned}
&f^{(n)}_0=\left.\frac{d^nf(r)}{dr^n}\right|_{r=r_0},\\
&f_0=f(r_0),
\end{aligned}
\end{equation}
we present a seven-term recurrence relation that reduces the time complexity to $\mathcal{O}(n)$:
\begin{equation}
\sum_{i=-4}^2a_i(n)R^{(n+i)}_0=0
\end{equation}
where
\begin{equation}
\begin{aligned}
a_{2}(n)=&\Delta_0^2,\\
a_{1}(n)=&2 \Delta _0 \left(r_0-1\right) (2 n+s+1),\\
a_{0}(n)=&2 n \left(a^2+3 r_0^2-6 r_0+2\right) (n+s)-(\Delta V)_0,\\
a_{-1}(n)=&2 n (n-1) \left(r_0-1\right) (2 n+3 s-1)-n(\Delta V)^{(1)}_0,\\
a_{-2}(n)=&n(n-1)\left[(n-2) (n+2 s-1)-\frac{(\Delta V)^{(2)}_0}{2}\right],\\
a_{-3}(n)=&2 \omega n(n-1)(n-2) \left(2 r_0 \omega +i s\right),\\
a_{-4}(n)=&\omega^2n(n-1)(n-2)(n-3).
\end{aligned}
\end{equation}
For higher-order derivatives of angular Teukolsky solutions $S_{lm}(c;x)$, with the definition Eq.~(\ref{eq: derivative notation}) but replace $r$ with $x$, there also exists a seven-term recurrence relation:
\begin{equation}
\sum_{i=-4}^2b_i(n)S^{(n+i)}_0=0
\end{equation}
where
\begin{equation}
\begin{aligned}
b_{2}(n)=&\vartheta_0^2,\\
b_{1}(n)=&-2 (2 n+1) x_0 \vartheta_0,\\
b_{0}(n)=&2 n^2 \left(3 x_0^2-1\right)+(\vartheta V_\theta)_0,\\
b_{-1}(n)=&2 n \left(2 n^2-3 n+1\right) x_0+n(\vartheta V_\theta)^{(1)}_0,\\
b_{-2}(n)=&(n-2) (n-1)^2 n-n(n-1)(\vartheta V_\theta)^{(2)}_0/2,\\
b_{-3}(n)=&2 c (n-2) (n-1) n \left(s-2 c x_0\right),\\
b_{-4}(n)=&-c^2 (n-3) (n-2) (n-1) n.
\end{aligned}
\end{equation}
with
\begin{equation}
\vartheta(x)=1-x^2
\end{equation}

However, the result we need is the minimal solution of the above recurrence relation.
\section{Check the Asymptotic Expansion at Infinity}\label{app: Check the Asymptotic Expansion at Infinity}
In Sec.~\ref{sec: Infinity Solution}, we give an asymptotic solution of Eq.~(\ref{eq: P(x)}) for large $x$; however, unfortunately, the convergence range is $0$. Although we can truncate at the ``best term" to achieve a high level of accuracy, the error term is highly undetermined. To prevent large accident errors from going undetected, the required function may need to be checked using another non-asymptotic method.

A consistent check was constructed using a method similar to the error estimation method in Eq.~(\ref{eq: residual equation}) of the main text. Assuming that an asymptotic series $\mathfrak{A}(x)$ can be used to obtain a rightness result at $x_0$, then it should satisfy Eq.~\ref{eq: P(x)}. Furthermore, to avoid the influence of the prefix exponential term (divergence at infinity), we perform additional transformations on Eq.~\ref{eq: P(x)}. In summary, we have
\begin{equation}
\begin{aligned}
(1-x) x& P_{\infty,\text{in}}''(x)+P_{\infty,\text{in}}'(x) q_3(x)+P_{\infty,\text{in}}(x) q_4(x)=0\\
q_3(x)=&-2 s x+s+i \tau -i \epsilon  (2 \kappa x (x-1)-2 x+3)-1\\
q_4(x)=&\lambda -\frac{(2 \epsilon -i) (\epsilon +i (s+i \tau -1))}{x}\\
&-i \epsilon  (\kappa -2 s+1)+2 s+2 \kappa  \epsilon ^2-2 \kappa  \tau  \epsilon
\end{aligned}
\end{equation}
\begin{equation}
\begin{aligned}
(1-x) x& P_{\infty,\text{out}}''(x)+P_{\infty,\text{out}}'(x) q_5(x)\\
+P_{\infty,\text{out}}&(x) q_6(x)=0\\
q_5(x)=&s (2x-3) +i(\epsilon+\tau) +2 i \epsilon \kappa  x^2 -2 i(\kappa +1) x \epsilon -1\\
q_6(x)=&\frac{(2 s+1) (s+1-i (\tau +\epsilon))}{x}\\
&+\lambda +i (\kappa +1) (2 s+1) \epsilon
\end{aligned}
\end{equation}
where
\begin{equation}
\begin{aligned}
P_{\infty,\text{in}}&=P_{\infty}^{\text{in}}/(-x)^{-i \epsilon+s+i \tau-1},\\
P_{\infty,\text{out}}&=P_{\infty}^{\text{out}}/e^{-2 i \epsilon \kappa x}/ (-x)^{i \epsilon-s+i \tau-1}.
\end{aligned}
\end{equation}
Then, we can define the residual $\varepsilon$ for $P_{\infty,\text{in}}$ and $P_{\infty,\text{out}}$. By checking whether $\varepsilon$ is smaller than $\bf{tol}$, the influence of the error term is tested. If the check result is false, then replace the valid radius $\mathcal{R}$ by $1.1~\mathcal{R}$. We will loop this procedure until the check result is true.

\section{Identity of the Wronskian Constant}\label{app: Identity of the Wronskians Constant}
We first give an explicit definition of the Wronskian constant $\mathcal{W}$ for an arbitrary differential equation
\begin{equation}\label{eq: arbitrary differential equation}
\frac{\text{d}^2w}{\text{d}x^2}+f(x)\frac{\text{d}w}{\text{d}x}+g(x)=0.
\end{equation}
The Wronskian constant of this differential equation is
\begin{equation}
\mathcal{W}(w_1,w_2)=\exp{\left[\int f(x) \text{d}x\right]}\mathcal{I}(w_1,w_2,x).
\end{equation}
This value depends on $w_1$ and $w_2$ and is independent of $x$. With this definition, the Wronskian constant of the Teukolsky equation in common sense is $\mathcal{W}(R^{\text{in}},R^{\text{up}})$.

We prove the following theorem:
\begin{theorem}
After transformation by any change of the dependent variable, take the following form:
\begin{equation}\label{eq: linear transform}
v(x)=t_1(x)w(x)+t_2(x)w^\prime(x)
\end{equation}
or any change of the independent variable $x=t_3(y)$, the Wronskian constant is identical to the previous one. We require functions $t_1$, $t_2$ and $t_3$ to be smooth.
\end{theorem}
We first deal with the change of the dependent variable. By adopting the transformation Eq.~(\ref{eq: linear transform}), the differential equation becomes
\begin{equation}\label{eq: transformed differential equation}
v^{\prime\prime}(x)+\left(f(x)-\frac{h^\prime(x)}{h(x)}\right)v^{\prime}(x)+s(x)v(x)=0
\end{equation}
where
\begin{equation}
h=t_1 \left(t_2^\prime-f  t_2\right)+t_2 \left(g t_2-t_1^\prime\right)+t_1^2,
\end{equation}
and $s(x)$ is a smooth function whose explicit form is useless for the following proof.
Thus, for the solutions $v_1$ and $v_2$ of Eq.~(\ref{eq: transformed differential equation}), the new Wronskian constant $\mathcal{W}_n$ reads
\begin{equation}
\begin{aligned}
\mathcal{W}_n(v_1,v_2)=&\mathcal{I}(v_1,v_2,x)\exp{\left[\int{\left(f(x)-\frac{h^\prime(x)}{h(x)}\right)}\text{d}x\right]}\\
=&h(x)\mathcal{I}(w_1,w_2,x)\exp{\left[\int{f(x)}\text{d}x-\ln{h(x)}\right]}\\
=&\mathcal{I}(w_1,w_2,x)\exp{\left[\int{f(x)}\text{d}x\right]}\\
=&\mathcal{W}(w_1,w_2)
\end{aligned}
\end{equation}
We then discuss the transformation of Eq.~(\ref{eq: arbitrary differential equation}) by a change of the independent variable. By denoting $w(t_3(y))$ as $u(y)$, after the transformation $x=t_3(y)$, the differential equation becomes
\begin{equation}
\begin{aligned}\label{eq: transformed differential equation2}
&u^{\prime\prime}(y)+\left((f\circ t_3)(y)t_3^\prime(y)-\frac{t_3^{\prime\prime}(y)}{t_3^\prime(y)}\right)u^\prime(y)+\\
&(g\circ t_3)(y)t_3^\prime(y)^2u(y)=0
\end{aligned}
\end{equation}
Thus, for the solutions $u_1$ and $u_2$ of Eq.~(\ref{eq: transformed differential equation2}), the new Wronskian constant $\mathcal{W}_n$ is
\begin{equation}
\begin{aligned}
&\mathcal{W}_n(u_1,u_2)\\
&=\mathcal{I}(u_1,u_2,y)\exp{\left[\int{\left((f\circ t_3)(y)t_3^\prime(y)-\frac{t_3^{\prime\prime}(y)}{t_3^\prime(y)}\right)}\text{d}y\right]}\\
&=t_3^\prime(y)\mathcal{I}(w_1,w_2,x)\exp{\left[\int{f(x)}\text{d}x-\ln{t_3^\prime(y)}\right]}\\
&=\mathcal{I}(w_1,w_2,x)\exp{\left[\int{f(x)}\text{d}x\right]}\\
&=\mathcal{W}(w_1,w_2)
\end{aligned}
\end{equation}

\section{Transition Points}\label{app: Transition Points}
The valid range of our expansion is determined by the transition points, i.e., the point at which $f(x)$ has a zero or singularity \cite{OLVER1974362}. In this appendix, we study the distribution of the transition points of Eq.~(\ref{eq: large omega f(x)}).

We consider the limit case $\omega\rightarrow\infty$. The zeros satisfy
\begin{equation}\label{eq: quartic equation}
\begin{aligned}
&x^4-\left(\frac{2}{\kappa}+2\right)x^3+\left(\frac{7+12\kappa}{4\kappa^2}+1\right)x^2\\
&-\frac{\kappa+4(1-\kappa)^2}{4\kappa^3}x+\frac{(1-\kappa)^2}{4\kappa^4}=0,
\end{aligned}
\end{equation}
which is a quartic equation and its roots can be represented in elementary functions.
\begin{figure}
\centering
\includegraphics[width=\linewidth]{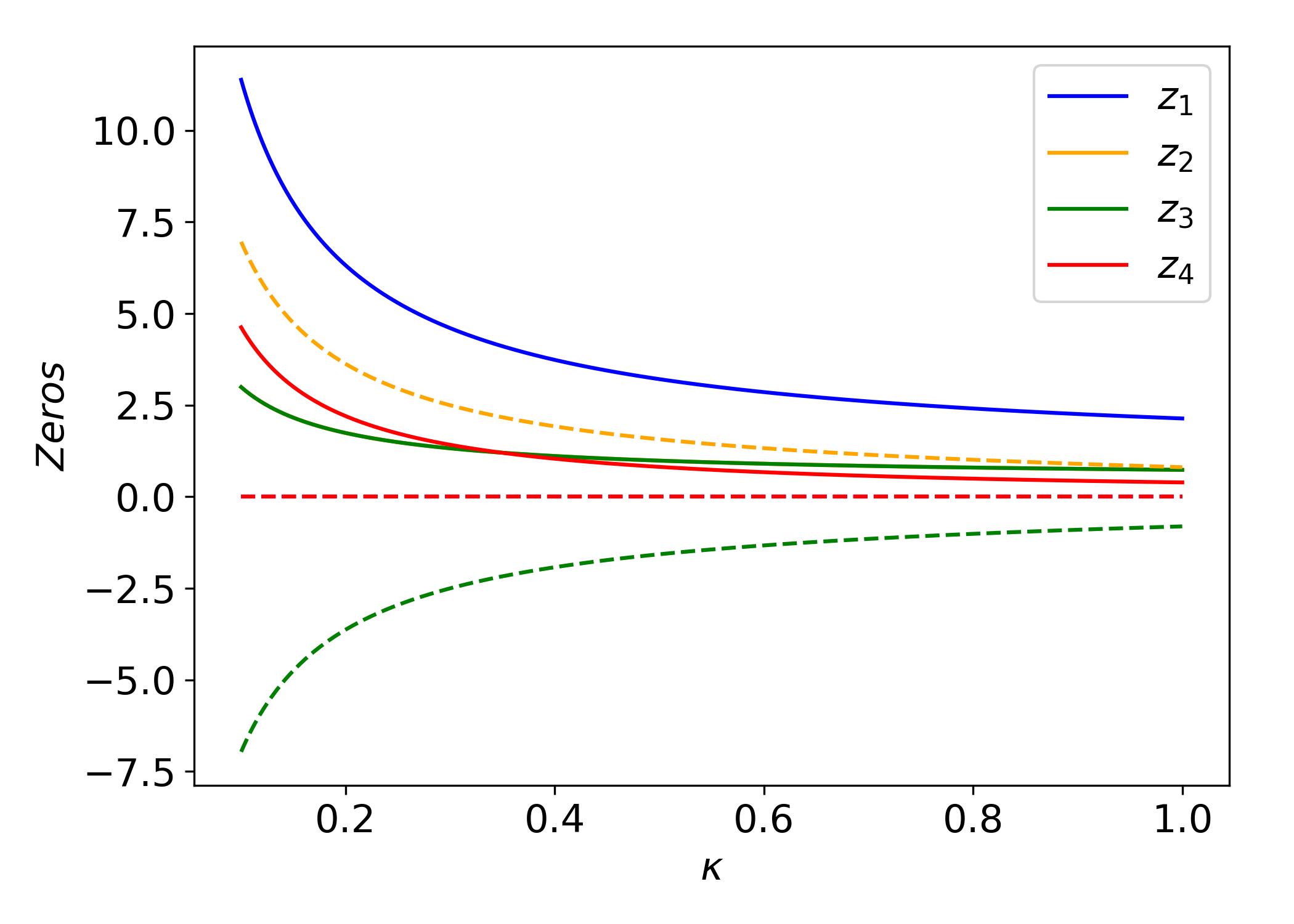}
\caption{Four solutions of Eq.~(\ref{eq: quartic equation}) respect to $\kappa$. The solid and dashed lines represent the real and imaginary parts, respectively. $z_1$ and $z_4$ are real solutions for $\kappa\in(0,1]$, $z_2$ is a conjecture with $z_3$.}
\label{fig: LOSE_root}
\end{figure}
Fig.~\ref{fig: LOSE_root} shows four roots of Eq.~(\ref{eq: quartic equation}) various with $\kappa$. These roots all lie in the right half complex plane; consequently, they cannot influence the convergence range of our expansion. 
\nocite{*}
%


\end{document}